\documentclass{article}

\usepackage{arxiv}

\usepackage[utf8]{inputenc} 
\usepackage[T1]{fontenc}    
\usepackage{hyperref}       
\usepackage{url}            
\usepackage{booktabs}       
\usepackage{amsfonts}       
\usepackage{nicefrac}       
\usepackage{microtype}      
\usepackage{lipsum}
\usepackage{graphicx}
\usepackage{amsmath}
\usepackage{subcaption}
\usepackage{multirow, makecell}
\usepackage{wrapfig}
\graphicspath{ {./images/} }

\title{Static IR Drop Prediction with Attention U-Net\\ and  Saliency-Based Explainability}

\author{
 Lizi Zhang \\
  University of Wisconsin-Madison\\
  Madison, WI 53706 \\
  \texttt{lzhang697@wisc.edu} \\
   \And
 Azadeh Davoodi \\
  University of Wisconsin-Madison\\
  Madison, WI 53706 \\
  \texttt{adavoodi@wisc.edu} \\
}

\begin{document}
\maketitle
\begin{abstract}
  There has been significant recent progress to reduce the computational effort of static IR drop analysis using neural networks, and modeling as an image-to-image translation task. A crucial issue is the lack of sufficient data from real industry designs to train these networks. Additionally, there is no methodology to explain a high-drop pixel in a predicted IR drop image to its specific root-causes. 

In this work, we first propose a U-Net neural network model with \textit{attention gates} which is specifically tailored to achieve fast and accurate image-based static IR drop prediction. Attention gates allow selective emphasis on relevant parts of the input data without supervision which is desired because of the often sparse nature of the IR drop map. We propose a two-phase training process which utilizes a mix of artificially-generated data and a limited number of points from real designs.  The results are, on-average, 18\% (53\%) better in MAE and 14\% (113\%) in F1 score compared to the winner of the ICCAD 2023 contest (and U-Net only \cite{UNet_Sachin_2021}) when tested on real designs. Second, we propose a fast method using \textit{saliency maps} which can explain a predicted IR drop in terms of specific input pixels contributing the most to a drop. 
    In our experiments, we show the number of high IR drop pixels can be reduced on-average by 18\% by mimicking upsize of a tiny portion of PDN's resistive edges. 
\end{abstract}


\section{Introduction}\label{sec:intro}

Static IR drop analysis of power delivery network (PDN) is a crucial task to accelerate IC design closure in modern technology nodes. 
Today's PDN is a large-sized 3D resistive network spanning across many metal layers. Due to parasitics in the PDN, voltage drop is induced between the power pads and cells in the design. A common goal of the analysis is to identify the nodes of the PDN where the voltage drop is higher than an acceptable threshold which  is typically referred to as the `hotspots'. 
Fast IR drop analysis is crucial to identify and guide many rounds of optimizations often needed to remove the hotspots, and accelerate design closure. 

Static IR drop analysis depends on: (1) topology of the PDN and resistance value of each branch in the network; (2) location of power pads on the PDN; (3) location and amount of current sources representing the cells or modules in the design. The underlying computation is based on solving a system of linear equations with millions to billions of variables \cite{Multigrid_Kozhaya_2002,FastAlg_Zhong_2005}.  A commercial tool  can take several hours to perform one round of analysis for modern  designs.

Earlier work for PDN analysis trade off accuracy for speed by utilizing spatial locality \cite{locality}, hierarchical methods \cite{Hierarchical}, Krylov-subspace methods \cite{Krylov_Hao_2001, Krylov_Chou_2011}, and multi-grid techniques \cite{Multigrid_Kozhaya_2002}. 
More recently, machine learning (ML) -based techniques provide significantly faster and more accurate solutions \cite{ECO-based_1, ECO-based_2, PowerNet}. But some are limited to incremental analysis and require retraining for each new design \cite{ECO-based_1, ECO-based_2}.  The work \cite{PowerNet} is based on Convolutional Neural Networks (CNNs)  
but assumes similar PDN resistance from each cell to  power pads, and configuring sizes of tiles on the layout. 

Most recently, \cite{UNet_Sachin_2021} models the problem as an image-to-image translation task. Each of the three inputs is expressed as an image. A U-Net model which is a convolutional encoder-decoder NN is used to generate a predicted IR drop map. This model naturally handles 2D spatially distributed data without the need to set any tile size, and does not require a retraining for each design. 

Despite the impressive success of ML-based approaches, an important, unresolved challenge is lack of sufficient data from real  designs to train the NNs. Recently, \cite{BeGAN_Chhabria_2021} created a large training dataset for this purpose. It utilized adversarial models to artificially generate data which had realistic current maps. Open-source static IR drop solvers were then used to generate golden ground-truth. 
However, there has not been a systematic evaluation of how NNs trained with artificially-generated data perform on real designs. The  ICCAD 2023 contest featured a problem on this issue \cite{ICCAD2023-contest}.

Finally, an issue with  \cite{UNet_Sachin_2021} is lack of explainability of a predicted hotspot to its root-causes. As an example, consider that the output generated by \cite{UNet_Sachin_2021} is an (2D) image but the PDN is a 3D structure. 
Predicting hotspots at specific pixels in a 2D image cannot be traced back to specific edges and layers. Neither can the hotspots be traced  to other sources such as specific locations on an input current density map or specific power pads. The degree of contribution of each of these cannot be determined. This explainability is a highly desired next step to guide optimizations to reduce the hotspots.

\textbf{In this work}, to address the above issues, we first propose \texttt{AttUNet}, a neural network model based on U-Net embedded with \textit{attention gates}. Attention gates allow  selective emphasis on relevant parts of the input data without supervision. This allows the model to focus on useful information during the learning process. Use of attention gates is motivated by the often sparse nature of the IR drop maps. In addition, the U-Net architecture is designed for the task of single-image to single-image prediction. However, IR drop is a multi-image to single-image prediction task. To handle this issue, we also embed the U-Net architecture with a new preprocessing convolutional block which introduces an initial per-image filter.


Next, we leverage a \textit{pretrain-finetune} strategy. We first train \texttt{AttUNet} using a large volume of artificially-generated data. A few data points from real design are then used in a final fine-tuning. We apply image transformation to augment the training data to improve the robustness of the training.

Finally, we propose a procedure to generate \textit{saliency maps} \cite{Saliency_Simonyan_2014} for a predicted IR drop map. These maps allow measuring degree of contribution of each pixel of each input image to any pixel in the output image. They are found via backward propagation of the output image and computing gradients with respect to each input pixel. A main advantage of saliency maps is the very low computational effort to generate them (e.g., seconds) without the need to modify the model. In contrast,  existing techniques for explainability of NNs often require modifying the model, e.g., by adding a new layer which compromises the performance \cite{CNNExplainer_Zhang_2018, XCNN_Tabanaei_2020, DAttn_Seo_2017}.

 The summary of our contributions is listed below:
\begin{itemize}
    \item We propose \texttt{AttUNet}, a new variation of UNet with embedded attention gates, tailored for the IR drop problem which is by nature a multi-image to single-image prediction task.
  \item We adapt a `pretrain-finetune' transfer learning strategy which is able to avoid overfitting and thus handle the issue of lack of sufficient data from real designs. We apply a data augmentation step to improve the robustness of training.
  \item We propose a procedure to generate saliency maps for a predicted IR drop image which allows measuring the contribution of each input feature to predicted high-drop pixels. 
\end{itemize}

In our simulations we use the ICCAD 2023 contest setup which provides 
a hundred of artificially-generated data points and few points of real designs. When measuring performance on real designs only, we show \texttt{AttUNet} is on-average 18\% (53\%) better in MAE and 14\% (113\%) in F1 score compared to the winner of the recent ICCAD 2023 contest (and U-Net only \cite{UNet_Sachin_2021}). We also use saliency maps for predicted high-drop pixels to guide PDN optimization. We show the number of predicted high IR drop pixels can be reduced on-average by 18\% by upsizing a tiny portion of resistive edges of the PDN. 

\begin{figure*}[]
    \centering
    \includegraphics
    [width=6in]{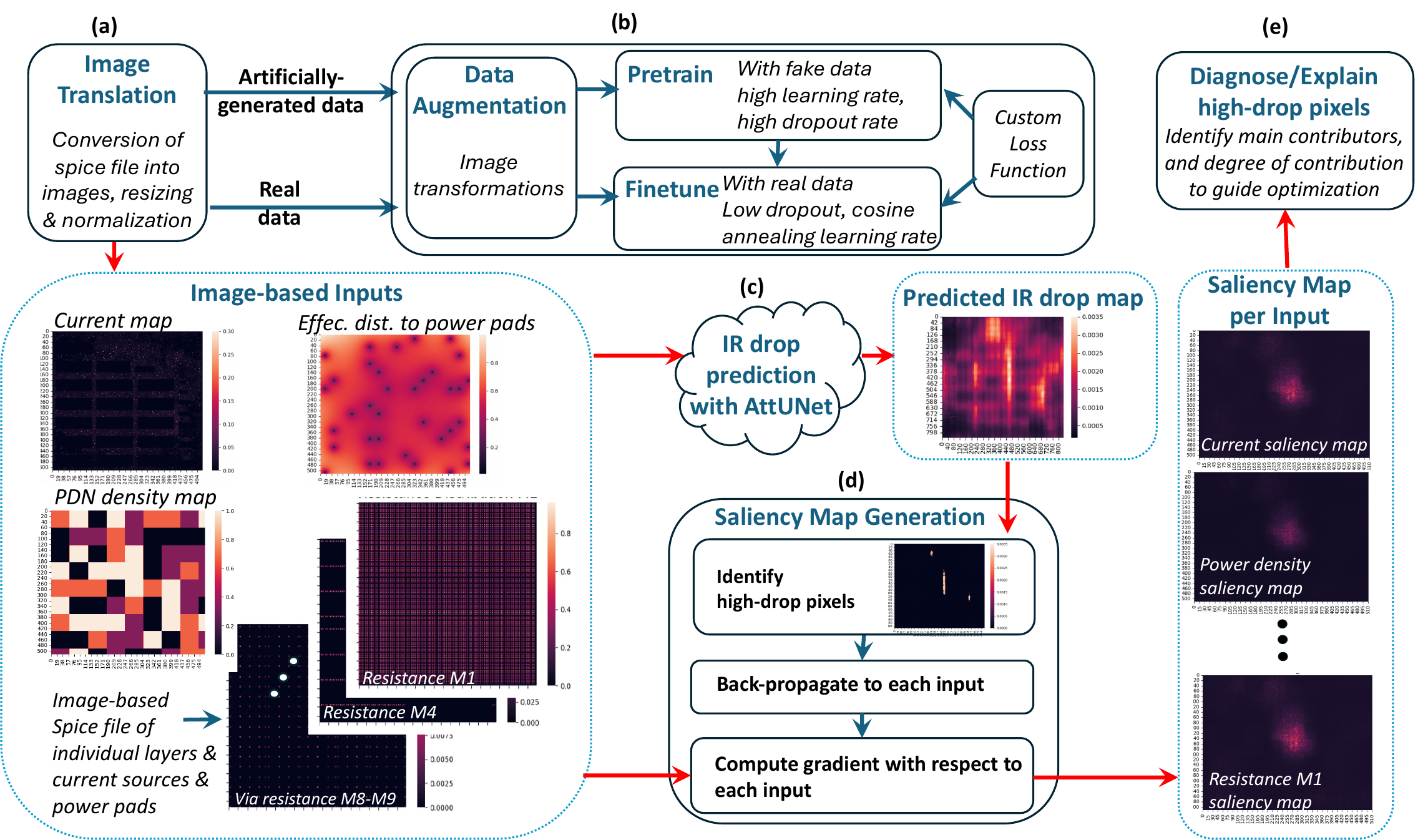}
    \caption{Components of our methodology: (a) image translation; (b) data augmentation and two-step training; (c) inference with attention U-Net; (d) saliency map generation; (e) explaintion and diagnosis of predicted hotspots.}
    \label{fig:flow}
\end{figure*}
\section{Proposed Methodology}\label{sec:method}
Figure \ref{fig:flow} shows the overall flow of our approach. The inputs to the problem are (1) current map which has the distribution of current sources (representing modules/cells) in the design; (2) PDN density map which includes the topology of the PDN and resistance values of each connection (via/metal layer); (3) map of effective distance to power pads which has the locations of all voltage sources (power pads) in the design. 

The \texttt{AttUNet} neural network receives these inputs and generates a predicted IR drop map as an image-to-image translation task. The first step shown in Figure \ref{fig:flow}(a) is to represent the inputs as images. We use 3 images representing the 3 inputs to the problem as in \cite{UNet_Sachin_2021}. Additionally, we translate the spice file describing all the layers into multiple images corresponding to each individual layer. 
These are explained in Section \ref{sec:image-prep}. Next, the \texttt{AttUNet} model is trained using the two-step pretrain-finetune strategy which is shown in Figure \ref{fig:flow}(b). Before training, we first augment input images by applying image transformations to increase the training size which helps improve the robustness of the trained model. The details are explained in Section \ref{sec:train}. Image-based inputs are then fed to the trained model for IR drop prediction as shown in Figure \ref{fig:flow}(b). We discuss the \texttt{AttUNet} model for IR drop prediction in Section \ref{sec:attunet}. 

After generating an IR drop map, saliency maps are generated as shown in Figure \ref{fig:flow}(d) and discussed in Section  \ref{sec:saliency}. Here, first high-drop pixels are identified from the generated IR drop map. The output of this step are individual saliency maps (one per input image). These saliency maps can be directly compared to each other because they are generated from normalized image-based inputs. They highlight degree of contribution of individual pixels of each input image to identified high-drop pixels which can be used to guide optimizations in order to generate a \textit{cooler} IR drop map image.

\subsection{Image-based Inputs}\label{sec:image-prep} 
The first step prior to inference or training is to translate the inputs to image-based format.  We explain how this is done including resizing and normalization operations relative to the setup in the recent ICCAD 2023 contest \cite{ICCAD2023-contest} which we used in our experiments. For each design, 3 image-based and 1 spice file describe the problem inputs. The 3 image-based inputs are described in  detail in \cite{UNet_Sachin_2021, ICCAD2023-contest}.


They correspond to the current map, PDN density map, and effective distance to power pads. 
(Current map shows locations of current sources.  PDN density map reflects the spacing between power stripes per unit area across all metal layers. Map of effective distance to power pads reflects the distance of each PDN node to all power pads, given by $\sum_{i=1}^N d_i^{-1}$ where $N$ is the number of pads and $d_i$  is distance to the $i^{th}$  pad.) These 3 inputs are expressed in matrix format where each entry in a respective matrix represents a pixel for an image-based representation. An entry in any of these matrices represents the value of the corresponding entity in the lowermost metal layer in $1\mu m\times 1\mu m$ area of the chip. 

\vspace{1mm}
{\noindent \textit{Converting Spice File into Additional Image-Based Inputs:}} In addition to the above 3 image-based inputs, in this work, a spice file is  added to describe detailed information across each individual metal and via layer. It contains  locations of PDN node, value of resistances between the nodes,  current source nodes and their values, and voltage source nodes. We extract multiple image-based files from this single spice file, where each new file correspond to the data of a specific metal or via layer. These are encoded in a similar matrix format. Each entry (pixel) represents a lumped resistance in a $1\mu m \times 1\mu m$ area. Specifically, for the ICCAD 2023 contest setup, we extract image-based files corresponding to resistances of layers M1, M4, M7, M8, M9, and via layers M14, M47, M78,  M89.

\vspace{1mm}
\noindent{\textit{\textbf{Resizing} and Normalization:}} Since the chip dimensions may be different, next we apply resizing to adjust all image-based inputs to the same dimension to allow processing by the same NN model. In general chip dimensions may vary among testcases. For example, in the ICCAD 2023 contest, they range from $80\times80$ to $1000\times1000$. In our implementation, and for this range, we resized all inputs to $512\times 512$ to feed the neural network. (This is in part because the encoder in the \texttt{AttUNet} model downsizes the images by power of 2 at each level.) For dimensions below $512\times 512$ upsizing is applied. It is done in a similar way using interpolation to fill in the `new pixels'. Anti-aliasing is enabled while resizing images to alleviate distortions. Typically, the distortions resulting from this resizing process are minor and have minimal impact on the performance of the model\footnote{We note, before conducting any evaluations in our experiments, we first resize each predicted IR drop map back to its original dimension to compare with a ground-truth.}.  Finally, for better adaptability, each input image is scaled to [0, 1] by dividing by its maximum matrix entry.

\subsection{U-Net Shaped Network with Attention Gate}\label{sec:attunet}
\begin{figure}[!t]
    \centering
    \includegraphics[width=0.75\linewidth]{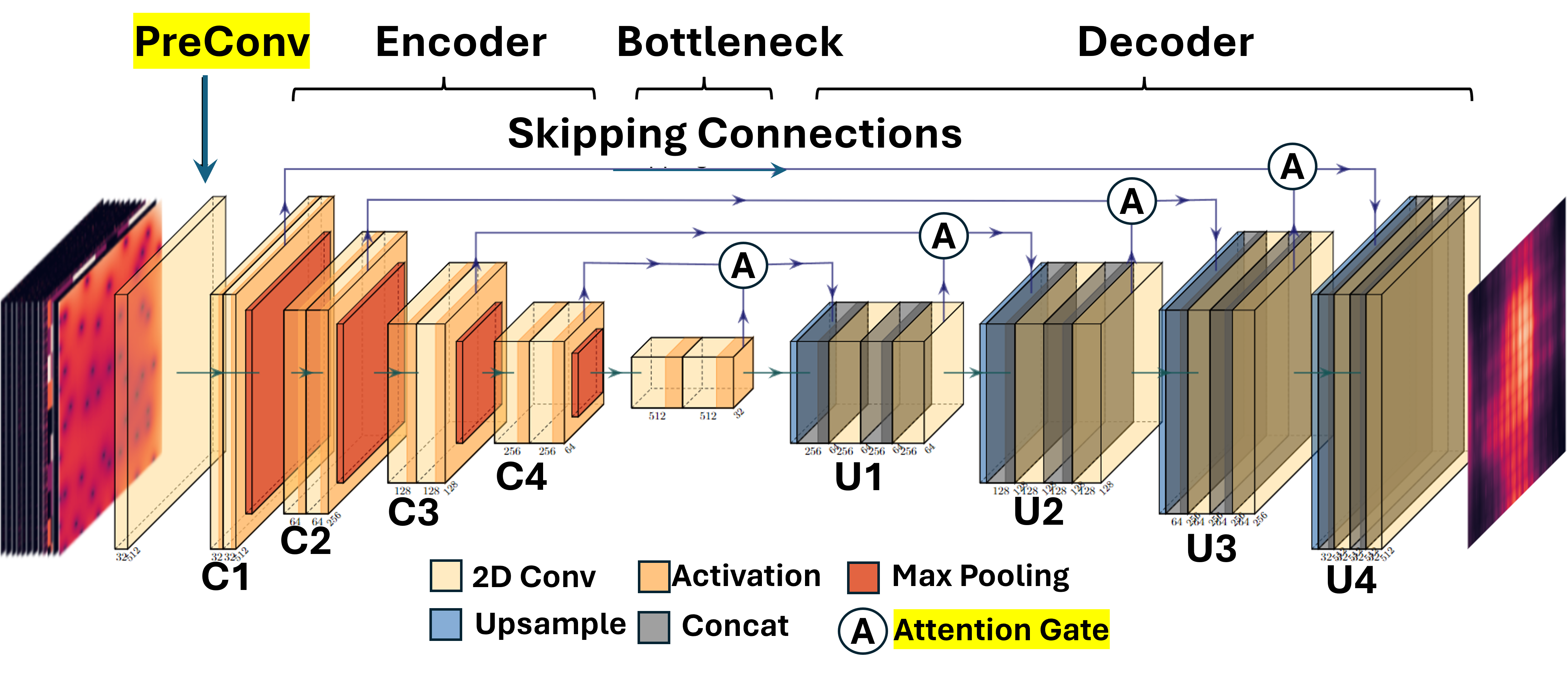}
    \caption{Block diagram of \texttt{AttUNet}. Additional embedded components compared to a U-Net structure are highlighted. }
    \label{fig:attunet}
\end{figure}

We give a brief overview of the U-Net architecture which was also used in \cite{UNet_Sachin_2021} for IR drop prediction. We discuss its limitations to motivate for our new embedded blocks. Figure \ref{fig:attunet} shows structure of \texttt{AttUNet}. Added blocks compared to U-Net are highlighted.

U-Net structure is well known for its great performance in image segmentation tasks. It has the ability to extract and exploit features to generate an image-based output. A U-Net consists of 4 major parts: encoder, bottleneck, decoder and skipping connection. Encoder, also called downsampling path, utilizes a sequence of double-convolutional blocks, each followed by batch normalization, a rectified linear unit (ReLU) and a max pooling layer. The bottleneck connects the encoder to the decoder. It is a double-convolutional block that receives feature maps which have a high number of channels (i.e., 512 in our implementation). 

The decoder is also composed of double-convolutional blocks, but instead of having a max pooling layer at the end, it has an upsampling layer at its beginning. A skipping connection exists between each encoder and decoder to pass global information of inputs to layers that are closer to outputs. Skipping connections allow the model to have a high-level view of the features and corresponding location information. The final layer is a $1\times1$ convolution that is used to map features to the desired number of classes.

There are several limitations of the U-Net model that make it less effective when dealing with IR-drop prediction. Firstly, it is originally designed for problems with single-image input. (The input is one image with multiple channels.)
In double-convolutional blocks, all channels are added together to generate new feature maps, which is effective for single-image problem because these channels have great similarity among each other and share the same features such as corners, edges, etc.
 However, for IR-drop prediction, the inputs to the model are multiple images and the images are very different from each other. 
Addition across channels entangles different features together and extracts inappropriate features by the model which can degrade the prediction performance. 

Moreover, U-Nets works best for image labeling and image segmentation, where the outputs are relatively simple (e.g., a single value). In contrast, in IR-drop prediction, a value should be predicted for each pixel in the output image and for the full-chip scale which is normally more than 10,000 pixels. 

To overcome these limitations, we propose \texttt{AttUNet} which is a new U-Net shaped neural network with new embedded components so it is better tailored for the multi-image-to-single-image translation tasks such as IR-drop prediction. The new components are highlighted in Figure \ref{fig:attunet}. We discuss each below.

\vspace{1mm}
\noindent{\textit{PreConv Block:}} The \texttt{AttUnet} network has an additional \texttt{PreConv} block inserted as the first layer to preprocess the inputs. The \texttt{PreConv} is a convolutional layer which modifies the inputs channel-wise before passing them to the encoder. 
It has a single convolution block with $2\times2$ filter and an activation function ReLU for each image-based input. The PreConv block works as a quick filter to highlight the salient features of each image-based input. Moreover, the convolution kernels (filters) scale each channel by assigning distinct weights to them, which reflect the overall significance of each channel. This preprocessing layer allows further convolution operations in U-Net to work with more relevant and distinguishing features, and better handle IR drop prediction as a multi-image to single-image task. 

\noindent{\textit{Attention Gates:}} These gates are added to the skipping connections between each pair of encoder and decoder in Figure \ref{fig:attunet}. They function as auxiliary components that selectively emphasize or suppress specific regions of the input data without supervision, enabling the model to concentrate on pertinent information throughout the learning phase \cite{AttUNet}. Given that the image-based inputs to IR drop problem are often sparse matrices (e.g., large portion of current maps are nearly zero), such regions have minimal contribution to IR-drop and can be suppressed by attention gates. 

\begin{figure}[h]
    \centering
    \includegraphics[width=0.75\linewidth]{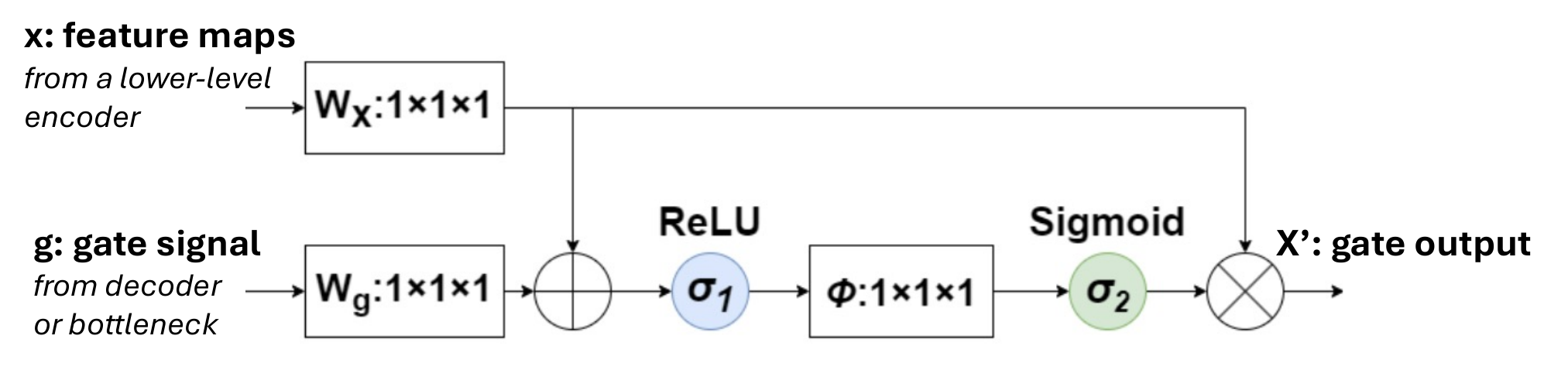}
    \caption{Vector concatenation-based attention gate.}
    \label{fig:attention_gate}
\end{figure}

Our work uses the vector concatenation-based attention gate proposed in \cite{AttUNet, Wang_NonLocalNN_2017} as shown in Figure \ref{fig:attention_gate}. The two inputs are  feature maps coming from a lower-level encoder, and a gate signal (which are feature maps coming from either bottlneck or decoder).

Denote $x_i\in\mathbb{R}^{C_x}$ and $g_i\in\mathbb{R}^{C_g}$ to be the vector elements of input feature maps and gate signal, respectively. The attention gate function can then be expressed as follows:
\begin{align}
\alpha_i&=\sigma_2(\phi^\top(\sigma_1(W_x^\top x_i+W_g^\top g_i+b_g))+b_\phi)\\
    x^\prime_i &=x_i\cdot \alpha_i
\end{align}
 The linear transformations are $W_x\in\mathbb{R}^{C_x\times C_l}, W_g\in\mathbb{R}^{C_g\times C_l}, \phi\in\mathbb{R}^{C_l\times 1}$ and the bias terms are $b_g\in\mathbb{R}^{C_l}, b_\phi\in\mathbb{R}$, and $\sigma _1$ is ReLU activation  and $\sigma_2$ is the sigmoid activation function. 
 
 The output is element-wise multiplication of $x_i$ and $\alpha_i$. Computation of $\alpha_i$ in (1) can be further expressed as:
\begin{equation}
    \alpha_i=\sigma_2(\phi^\top(\sigma_1(W^\top [x_i,g_i]+b))+b_\phi)
\end{equation}
where $W\in \mathbb{R}^{(C_x+C_g)\times C_l}$ is concatenation of $W_x$ and $W_g$ along the second dimension. $[x_i,g_i]$ is a vector concatenation of $x_i$ and $g_i$. This operation is also the reason why this attention gate is called vector concatenation-based attention gate.

With expression in (3), the linear transformation can be easily done by using convolutions directly on gate inputs with $1\times1$ kernel and $C_{l}$ output channels. Then, the processed feature map is activated by a ReLU function and is linearly mapped to a $\mathbb{R}^{C_l}$ space. The output of sigmoid activation function $\alpha\in (0,1)$ is known as attention map (or coefficients). These coefficients assign weights to each value of input feature maps coming from the encoder. The attention gate also receives the gate signals from deeper layers (e.g., bottleneck or decoder) to suppress irrelevant information and noise as they go through a skipping connection.

\subsection{Data Augmentation and Model Training}\label{sec:train}
We use a `pretrain-finetune' strategy for training \texttt{AttUNet} as shown in Figure \ref{fig:flow}(b). First, pretrain is done using large volume of artificially-generated data. Next, finetune is done using limited data from real designs. (The provided dataset in the ICCAD 2023 contest contains 120 test cases in total, of which 100 are artificially-generated explained in \cite{BeGAN_Chhabria_2021,ICCAD2023-contest}, and the remaining 20 are real. ) Each test case is represented as image-based inputs as discussed in Section \ref{sec:image-prep}, and  is also accompanied by an image-based golden output voltage file.

\begin{figure}
    \centering
    \includegraphics[width=0.6\linewidth]{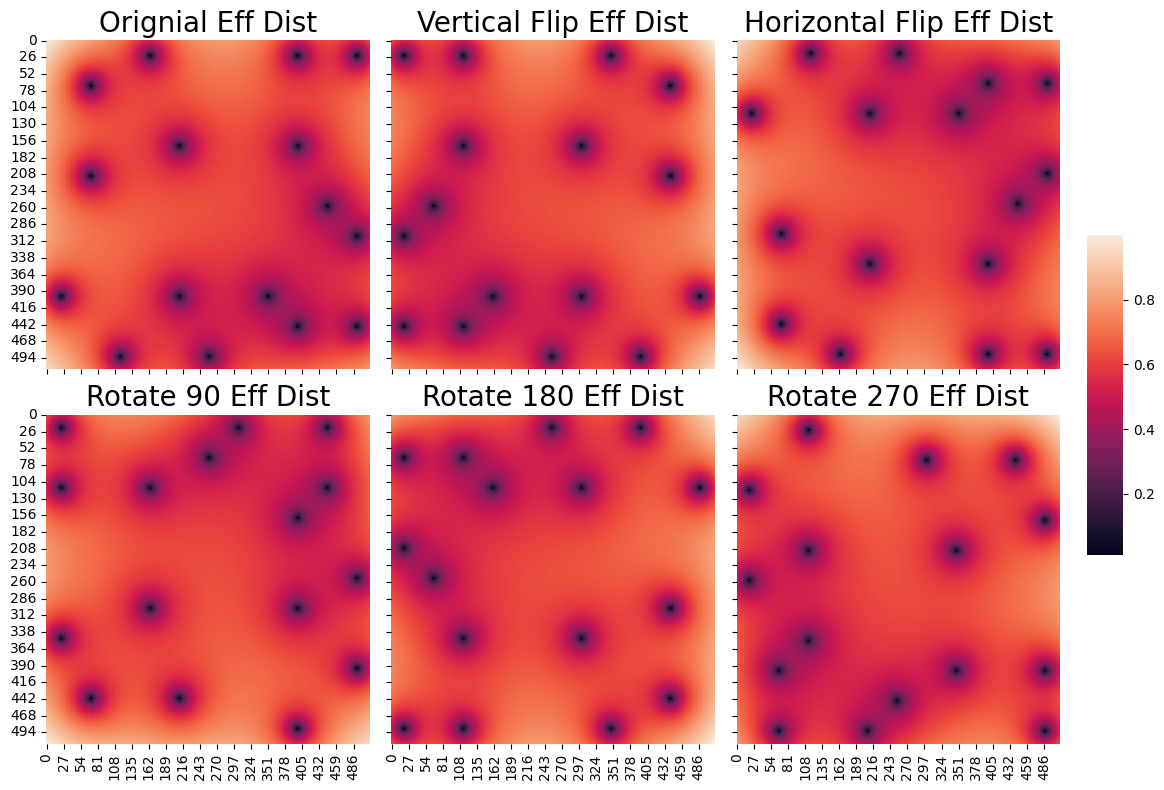}
    \caption{Augmenting the training data via five transformations applied to each image-based input.}
    \label{fig:data_augmentation}
\end{figure}

{\noindent \textit{Data Augmentation:}} As shown in Figure \ref{fig:flow}(b), we first augment the training data by applying multiple transformations to each image-based input which help improve the robustness of the model \cite{ImageNet_Krizhevsku_2012} especially when training data is not sufficient. Specifically, we apply the following five operations to each image-based input: vertical and horizontal flipping and  three (counter-clockwise) rotations as shown in Figure \ref{fig:data_augmentation} for a sample effective distance map. Next, a new testcase is generated by applying one of the five operations to an existing testcase; For instance, vertically flipping all the image-based representations within a testcase generates a new augmented testcase. This process results in a sixfold increase in the number of testcases, and enhances the diversity and robustness of the dataset \cite{ImageNet_Krizhevsku_2012}. It is applied to both artificially-generate data in pretrain phase, as well as real data in finetune phase.

We note, horizontal and vertical flippings only rearrange the elements in matrix representation of each input. These do not alter the validity of the results when solving the linear system of equations describing the IR-drop problem. Similarly, rotation operations essentially prompt the network to view the images from different angles, ensuring the results remain valid. The rotation angles are restricted to $90^{\circ}, 180^{\circ}, 270^{\circ}$ to avoid cropping and centering operations which may change the resulting IR-drop.



{\noindent \textit{Model Training:}} To train \texttt{AttUNet}, a transfer learning approach is employed. 
First, the pretraining phase aims to maximize the utilization of generative fake data. (The fake data contains meaningful input-output information because according to \cite{BeGAN_Chhabria_2021}, an actual open-source IR drop solver is used to generate its golden ground-truth.) Next, finetuning is conducted only based on real data.

\begin{figure}
    \centering
    \includegraphics[width=0.75\linewidth]{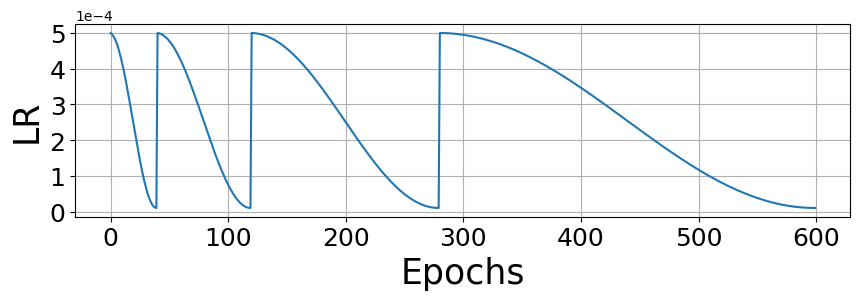}
    \caption{Illustration of the cosine annealing learning rate.}
    \label{fig:cos_anneal}
\end{figure}

The learning rate and drop rate of the pretrain and finetune phases are listed in Table \ref{tab:attunet}.  For the pretraining phase, we first use a relatively high learning rate of 0.0005 and a high dropout rate from 0.3 to 0.5.  For the fine-tuning phase which is based on real data, we adjust to a finer learning granularity, ranging from 0.0005 to 0.00001. Also, a cosine annealing learning rate with restart is applied in the finetune phase which can be expressed as follows:
\begin{equation}
    \eta_t = \eta_{min}+\frac{1}{2}(\eta_{max}-\eta_{min})(1+cos\frac{T_{cur}}{T_{max}}\pi)
\end{equation}
where $\eta_{max}$ is set to be the initial learning rate 0.0005, $\eta_{min}$ is the lowest learning rate 0.00001. The parameter $T_{cur}$ is the number of current epochs and $T_{max}$ is the number of total training epochs (600). As shown in Figure \ref{fig:cos_anneal}, learning rate oscillates with a decreasing frequency with increase in  epochs. This provides a balance between rapid exploration of the parameter space at higher learning rates and finer optimization in regions of interest at lower learning rates. 

Cosine annealing helps the model converge faster during the early stages of training  \cite{cosine_anneal}: By starting with a higher learning rate and gradually reducing it, the optimization process becomes more efficient. The cyclical nature of learning rate also allows the model to escape local minima to better explore the loss landscape.

As a final consideration during training, we employ a custom loss function. We note, the goal of IR-drop analysis is to predict the hotspot locations. Underestimating these is undesirable. Therefore, we define a custom loss function while training the model, which motivates the model to err on the side of overestimating the IR-drop, even if it results in a larger error. The loss function is set to punish more when a predicted value is less than the actual value. which can be expressed as 
\[
    Loss= 
\begin{cases}
    \frac{1}{n}\sum^{n}_{i=1}\|x_i-y_i\|,& \text{if } x_i\geq y_i\\
    \frac{\lambda}{n}\sum^{n}_{i=1}\|x_i-y_i\|, & \text{if} x_i<y_i
\end{cases}
\] 
where $\lambda\geq 1$ is a constant, $n$ is the number of pixels in the output figure, $x_i$ is the predicted value and $y_i$ is the ground truth. We set $\lambda=2$ in our experiments. 

\subsection{Fast Explainability with Saliency Maps}
\label{sec:saliency}

A predicted IR-drop map can be used to identify high-drop areas (which are individual pixels in the generated image-based output). A desired next step is to understand which specific inputs to the neural network are most responsible for these high-drop areas. This would be identifying specific image-based inputs, and specific pixels within each image since our problem is a multi-image to single-image prediction task. Among these identified input pixels, we would also be able to compare how much is the contribution of each one to the predicted high-drop pixels at the output. 

These diagnosis help guide necessary optimizations to create a `cooler' IR-drop map. For example, if specific pixels in the current map are found to have the highest contribution then it suggests that changing the floorplan of modules or placement of cells may be most appropriate to reduce the current demand at those locations. If the highest contributors are specific PDN edges on specific metal layers, then slight upsizing of these edges may be most appropriate. Such explainability has not been explored in any prior work for static IR-drop prediction and is a natural and important next step.


However, \texttt{AttUNet}, like most deep neural networks, operates more as a `black box' which makes it difficult to comprehend the reasoning behind having specific output predictions. Existing techniques for adding explainability to a deep neural network often require changing the network structure, for example by adding extra layer(s) which can in turn compromise the performance \cite{CNNExplainer_Zhang_2018, XCNN_Tabanaei_2020, DAttn_Seo_2017}. However, saliency maps are available tools which allow gaining some insights into model behavior very quickly (e.g., seconds in our problem). In particular, for our problem which is an image-to-image translation task, we show the insight gained by saliency maps can be helpful for diagnosing the predicted high-drop pixels.



Figure \ref{fig:flow}(d) shows the process of generating the saliency maps. The first step is identifying high-drop pixels from the predicted IR-drop map. This relies on the designer to know how much IR-drop is considered acceptable given the desired specifications. 
Next, a back-propagation is done from this \textit{high-drop-only} output map to each individual pixel on the input side. Finally, a gradient is computed with respect to each input. Since all image-based inputs are normalized to [0,1] range, the corresponding gradients are comparable across the pixels of different inputs. 
In the end a saliency map is generated for each image-based input, as shown in Figure \ref{fig:flow}(d).

Formally, let $F_k:\mathbb{R}^{C\times h\times w}\rightarrow \mathbb{R}$ denote the function describing how the model generates the $k^{th}$  pixel in the output image from the $C$ (single-channel) input images. Each image (inputs or the output) has a height $h$ and width $w$ in \texttt{AttUNet}. Let $F_k(X) = y_k$, $X\in \mathbb{R}^{C\times h\times w}$ denote all input images as a 3D matrix representing a collection of individual single-channel 2D inputs. Also $y_k\in \mathbb{R}$ is a pixel $k$ of the generated output image. 

Due to the complex nature of the neural network, $F_k$ is a highly non-linear function of $X$. However, given an input $X_0$, $F_k(X_0)$ can be approximated with a linear function in the neighborhood of $X_0$ by computing the first-order Taylor expansion \cite{Saliency_Simonyan_2014}: 
\[ F_k(X)\approx w^T X+b\]
where $w$ is the derivative of $F_k$ with respect to the input $X$ at $X_0$:\[w=\frac{\partial F_k(X)}{\partial X}\Bigg |_{X_0}\]

The magnitude of elements of $w$ defines the importance of the corresponding pixels of $X$ for the $k^{th}$ pixel in output image. 

The saliency map $S\in \mathbb{R}^{C\times h\times w}$ is computed in the similar way. For a subset of output pixels $\{y_k\}$, $k=1,2,\ldots,K$ (for example representing the predicted high-drop pixels), a (combined) saliency map $S$ is generated for input $X$ representing all image-based inputs. This is done by computing the average gradient regarding the subset of output pixels: \[ S=\frac{1}{K}\sum_{k=1}^K\frac{\partial F_k}{\partial X}\]
The above will have dimension $C\times h\times w$ and further be broken into individual saliency maps representing each input-based image.

\section{Experimental Results}
\begin{table}
    \centering
    \small
        \caption{Model hyperparameters and training parameters}
    \begin{tabular}{c|c|c|c}
    \toprule
        \multirow{11}*{\makecell{Model \\ hyperparameters}} &PreConv &filter size & 2×2 \\ 
         & & \# filters & 12\\ \cline{2-4}
           & C1 & filter size & 3×3\\
           & U1 & \# filters & 32\\ \cline{2-4}
           & C2 & filter size & 3×3\\
           & U2 & \# filters & 64\\ \cline{2-4}
           & C3 & filter size & 3×3\\
          &  U3 & \# filters & 128\\ \cline{2-4}
           & C4 & filter size & 3×3\\
           & U4 & \# filters & 256\\ \cline{2-4}
           & Bottleneck & filter size & 3×3\\
           &  & \# filters & 512\\ 
           \midrule
        \multirow{5}*{\makecell{Training\\ parameters}} &&Pre-train &Fine-tune\\ \cline{2-4}
         &Epochs& 450& 600\\\cline{2-4}
        &Optimizer& ADAM& ADAM\\\cline{2-4}
        &Learning rate& 0.0005& 0.00001-0.0005\\\cline{2-4}
        &Dropout& 0.3-0.5& 0.1\\
        \bottomrule
    \end{tabular}
    \label{tab:attunet}
\end{table}

The \texttt{AttUNet} model is implemented using Python 3.9 and under Pytorch 2.0.1 framework and is trained on two NVIDIA GeForce RTX 2080Ti GPUs. Table \ref{tab:attunet} describes the model hyperparameters and training parameters. Layer IDs are shown in Figure \ref{fig:attunet}.


We used the data provided by the ICCAD 2023 contest \cite{ICCAD2023-contest}. As mentioned earlier, 120 test cases were provided, 20 of which corresponded to real designs and the rest were artificially-generated based on \cite{BeGAN_Chhabria_2021}.  All 100 artificial data were used in the pre-train phase. From the 20 real designs, 10 were used in the fine-tune phase. The remaining 10 designs were used to test the model.  To be comparable to the contest results, we use the same train/test split as the contest, that is using the same ten real chip data as testing dataset, and the remaining data for training. 

As explained earlier, we pretrain \texttt{AttUNet} on augmented fake data and fine-tune on augmented real data. The augmentation enlarges the dataset by 6 folds, resulting in 800 pretrain data points and 80 fine-tune data points. The pretraining run time is 17.4 hours
and fine-tuning is 4.6 hours. It is important to note that this is a
one-time cost; the trained model can be applied directly to different
chip designs and does not need any  modifications. The average inference
time across the designs in the testing dataset is 5.37 seconds.

We compare \texttt{AttUNet} with the  UNet model in \cite{UNet_Sachin_2021} which is named as \texttt{IREDGe}. We implemented this model based on the   hyperparameters in \cite{UNet_Sachin_2021}. For a fair comparison using the contest setup, we trained \texttt{IREDGe} using the same pretrain and fine-tune stages and same training data. For \texttt{IREDGe}, we used all training parameters reported in \cite{UNet_Sachin_2021} except the number of epochs and learning rate. These two were set the same as \texttt{AttUNet} which generated better results. We also added the same dropout as \texttt{AttUNet}.

Additionally, we compare with the winner of the ICCAD 2023 contest by directly using the quality metrics reported by the contest for the winning team. According to the contest website, the winning team uses a modified version of the \texttt{ConvNetXTV2}.




To measure quality of prediction, we use the same metrics given in the contest. These are (1) Mean Absolute Error (MAE) and (2) F1 score. First, MAE is computed as the absolute error between predicted IR-drop and the ground-truth given by:  \[MAE=\frac{\sum_i^N{|\hat{V_i}-V_i|}}{N}\] where $\hat{V}$ is  predicted IR drop, $V$ is ground truth and $N$ is number of pixels. This metric reflects an \textit{overall} prediction accuracy. 

To compute an F1 score, first the highest 10\% IR drops are labeled as positive and the rest as negative. The F1 score is given by:
\[F1 = \frac{2\times precision\cdot recall}{precision+recall}\]
\[ Recall = \frac{\text{True Positives}}{\text{True Positives + False Negatives}} \]
\[ Precision = \frac{\text{True Positives}}{\text{True Positives + False Positives}} \]
where precision is the ratio of true positive predictions to the total number of predicted positives, and recall is the ratio of true positive predictions to the total number of actual positives. F1 score assesses the trade-off between precision and recall. 


\subsection{Prediction Quality}

Table \ref{tab:prediction quality} compares our F1 score and MAE to \cite{UNet_Sachin_2021} and the contest winner. The bold entries indicate the best MAE and F1 score in each row.
First, \texttt{AttUNet} significantly outperforms \texttt{IREDGe} (on-average 113.3\% in F1 score and 52.7\% in MAE) under the contest setup. Furthermore, \texttt{AttUNet} outperforms \texttt{ConvNeXtV2} in many cases. It is better in 9 out of 10 designs in at least one metric (MAE or F1 score). It achieves the lowest average MAE (0.110 mV) and highest average F1 score (0.520) among all test cases which is about 18.1\% improvement in MAE and 14.3\% in F1 score compared to \texttt{ConvNeXtV2}.  Notable improvements are in test cases 13 and 14, in which \texttt{ConvNeXtV2} has a 0.000 F1 score, indicating an inability to highlight the severe IR drop regions. \texttt{AttUNet} has F1 scores of 0.895 and 0.881 for these. 
\begin{table}[!t]
\small
    \centering
        \caption{Comparison of IR-drop prediction quality using the ICCAD 2023 contest setup. The unit for MAE is mV. Lower MAE and higher F1 score are  desired.}
    \begin{tabular}{c|cc|cc|cc}
    \toprule
    &\multicolumn{2}{c|}{\texttt{IREDGe} }  & \multicolumn{2}{c|}{\texttt{ConvNeXtV2}}&\multicolumn{2}{c}{\texttt{AttUNet}} \\
        &\multicolumn{2}{c|}{\cite{UNet_Sachin_2021}}  & \multicolumn{2}{c|}{(Contest Winner)}&\multicolumn{2}{c}{} \\
    \midrule
    & MAE & F1 & MAE & F1 & MAE & F1 \\
    \midrule
    Testcase7 &  0.124 & 0.648 & 0.066 & \textbf{0.783} & \textbf{0.065} & 0.469\\
    Testcase8 &  0.110 & 0.698 & 0.082 & \textbf{0.816} & \textbf{0.081} & 0.428\\
    Testcase9 &  0.205 & 0.120 & \textbf{0.041} & \textbf{0.589} & 0.067 & 0.169\\
    Testcase10 & 0.141 & 0.483 & \textbf{0.066} & 0.532 & 0.090 & \textbf{0.536}\\
    Testcase13 & 0.119 & 0.417 & 0.207 & 0.000 & \textbf{0.188} & \textbf{0.895}\\
    Testcase14 & 0.192 & 0.034 & 0.422 & 0.000 & \textbf{0.078} & \textbf{0.881}\\
    Testcase15 & 0.157 & 0.000 & \textbf{0.097} & 0.088 & 0.115 & \textbf{0.700}\\
    Testcase16 & 1.066 & 0.000 & \textbf{0.160} & 0.529 & 0.338 & \textbf{0.634}\\
    Testcase19 & 0.131 & 0.037 & 0.091 & \textbf{0.501} & \textbf{0.082} & 0.330\\
    Testcase20 & 0.089 & 0.000 & 0.118 & \textbf{0.711} & \textbf{0.068} & 0.156\\

    \midrule
    Average& 0.233 & 0.244 & 0.135 & 0.455 & \textbf{0.110}  & \textbf{0.520} \\
    \bottomrule 
    \end{tabular}

    \label{tab:prediction quality}
\end{table}

\begin{figure}[t]
    \centering
    \includegraphics[width=0.75\linewidth]{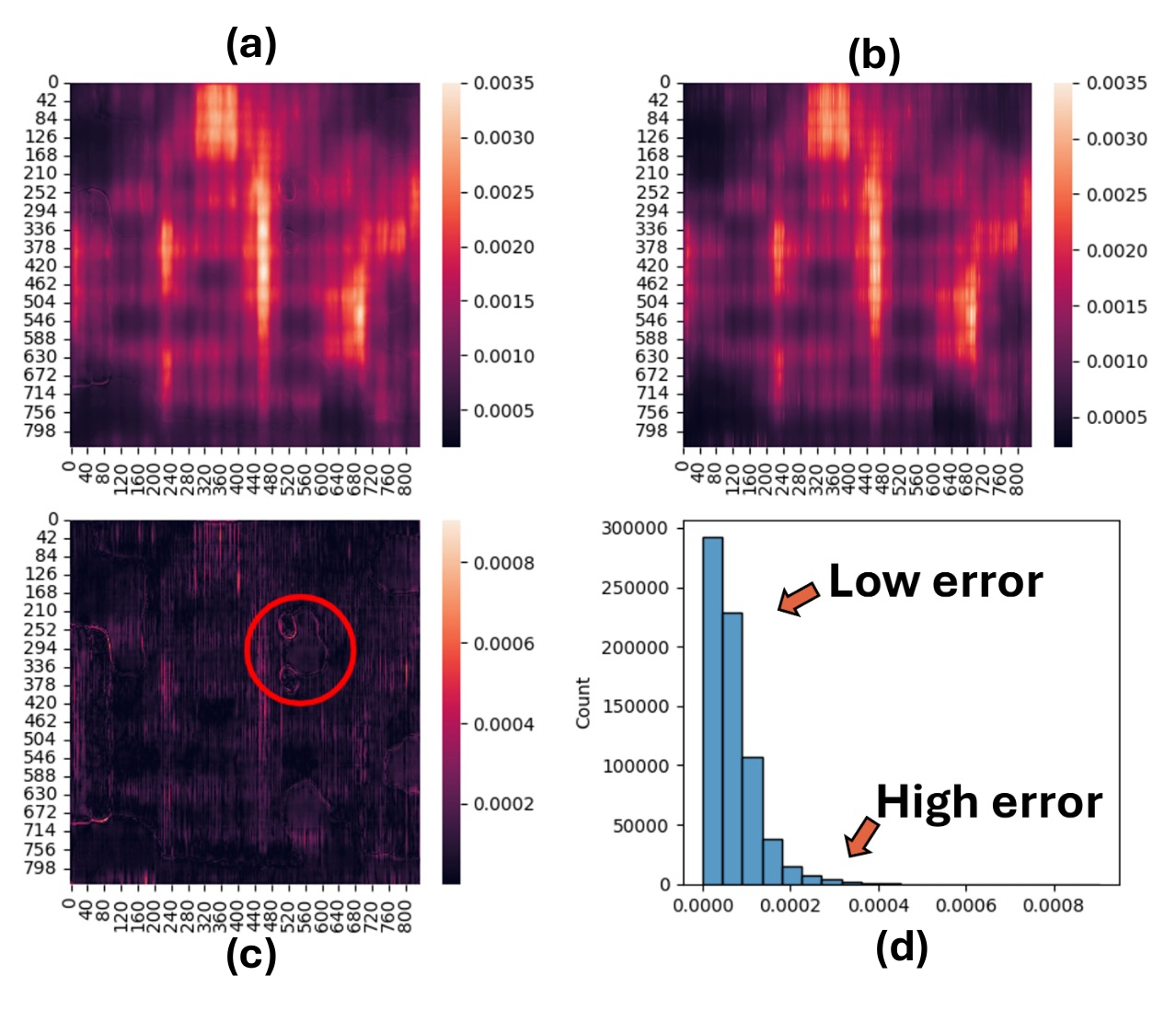}\vspace{-3mm}
    \caption{Results for testcase 9: (a) ground truth map, (b) map generated by \texttt{AttUNet}, (c) MAE map (d) Distribution of MAE. }\vspace{-5mm}
    \label{fig:prediction}
\end{figure}

Figure \ref{fig:prediction} shows the output of \texttt{AttUNet} in testcase 9. As shown in (a) and (b), our prediction closely aligns with the ground truth in spotting the high-drop areas. 
Figures (c) and (d) show the map and distribution of MAE. 
Location of the spots with higher prediction error can be seen in (c) such as the circled region. However, from the distribution we can see that most-frequent error bin (left-most bar in (d)) is about 0.2 mV. This value is significantly smaller than of the high error bins (which are higher than 3.0 mV). 

\begin{table}[t]
    \centering
    \caption{Comparison of the two metrics (averaged) shown for the training set (and  the testing set, in parenthesis).}
    \small
    \begin{tabular}{c|cc|cc}
    \toprule
     & \multicolumn{2}{c|}{Pre-trained Model}&\multicolumn{2}{c}{Fine-tuned Model}  \\
    \midrule
    & MAE (mV) & F1 Score& MAE (mV) & F1 Score\\
    \hline
    \texttt{IREDGe} & 0.23 (1.14) & 0.24 (0.17) &  0.13 (0.23) & 0.32 (0.24)\\
    \texttt{AttUNet} &  0.07 (0.38) & 0.57 (0.32) & 0.05 (0.11) & 0.54 (0.52) \\
    \bottomrule
    \end{tabular}
    \label{tab:ave_performance}
\end{table}

\subsection{Training Efficiency}

Table \ref{tab:ave_performance} shows the average training (testing) MAE and F1 scores of \texttt{IREDGe}, \texttt{AttUNet} at different training stages. After fine-tuning, \texttt{AttUNet} achieves an MAE of  0.0520 mV and F1 score of 0.541. The MAE is significantly smaller compared to IR-drop values which normally range from $1\textup{--}10$ mV. Compared to \texttt{IREDGe}, \texttt{AttUNet} has better training and testing performances in terms of both average MAE and F1 score, indicating the effectiveness of learning and fitting the patterns and relationships present in the training data.

Given the inadequacy of real data, pre-training with generative fake data serves as a preventive measure against the potential risk of over-fitting. Generative errors (gaps between training MAE and testing MAE) decreases from 0.31 to 0.06 after fine-tuning, showing a great reduction in over-fitting risks. Also, F1 score after fine-tuning is at the same level in training and testing datasets. In comparison, \texttt{IREDGe} shows much bigger generative errors, which are 0.91 after pre-training and 0.10 after fine-tuning, indicating a significant possibility of over-fitting.

\begin{figure}
    \centering
    \includegraphics[width=0.85\linewidth]{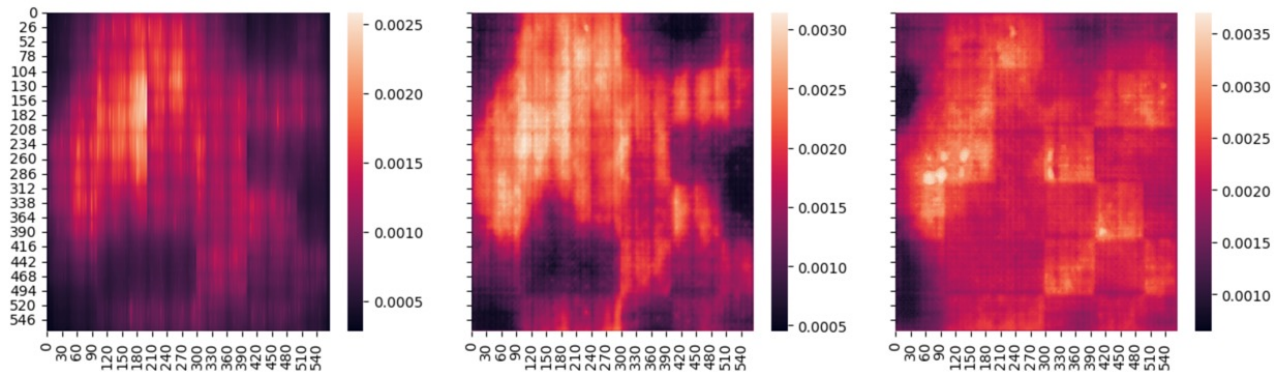}\vspace{-1mm}
    \caption{Impact of pre-training: Ground truth (left); \texttt{AttUNet} after pretraining and 50 epochs of fine-tunning (middle); same \texttt{AttUNet} but without pretraining (left).}
    \label{fig:pretrain}
\end{figure}
Figure \ref{fig:pretrain} shows the output of AttUNet in the early stage of fine-tuning, specifically after 50 epochs. When the model is pre-trained (middle figure), the output shows clear boundaries that successfully separate regions with high IR-drop. In contrast, the model without pre-training (shown in right) lacks these distinctive features.

\subsection{Explainability with Saliency Maps}
In this experiment, we first generate a saliency map for each image-based input. Recall, this requires identifying high-drop pixels from the predicted IR-drop image at the output. Specifically, for each testcase, we identify the maximum predicted voltage drop across all pixels which we denote by $Dr_{max}$. We then extract \textit{high-drop} output pixels as those with a drop higher than $Dr_{th}=0.9\times Dr_{max}$. 

Once saliency maps are obtained, we identify the top $K$ pixels from the input images which have the highest contribution to these high-drop output pixels. To identify the top $K$ input pixels, we first compute an average saliency for each input image by averaging the top $K$ pixels. The input image with the highest contribution is the one with highest average saliency. Next, within that image we identify the top $K$ pixels which have the highest saliency values.

Next we show how these top $K$ input pixels can guide optimization to generate a new IR drop map with fewer high-drop pixels. The optimization that we considered in this experiment is upsizing the PDN wire branches corresponding to the top $K$ input pixels.  Therefore, we only considered the saliency maps of 5 metal layers (M1, M4, M7, M8, M9) and 4 via layers (M14, M47, M7M, M89). Other optimizations such as change in floorplan was not possible due to lack of more detailed layout information. Also, our emphasis in this experiment is to only show the impact of saliency maps. Exploring the best optimization is outside the scope of this work.

Table \ref{tab:explainability} shows the results for $K=100,300,500$. We note these values of $K$ represent a tiny portion of each $512\times 512$ input image. Our optimization mimics  upsizing some PDN wire widths (thus reducing their resistances) which correspond to the top $K$ pixels. Upsizing is mimiced by reducing the corresponding pixel values (in the corresponding image) by 10\%. For example, in testcase 7 the top contributor layer is M1. It has 1574 high-drop pixels in the predicted output image. We identify only $K$ pixels in the image input corresponding to M1 for optimization. 


Next, we feed this slightly-optimized input back to \texttt{AttUNet} to predict a new IR-drop map.  Using the same $Dr_{th}$ as earlier, we then identify the number of high-drop pixels in the optimized map. We report percentage reduction in the number of high-drop pixels in Table \ref{tab:explainability}. On-average the number of predicted high-drop pixels are reduced by 18.48\%, 37.28\%, 49.05\%  for $k=100,300,500$, respectively. 

To show identifying these top $K$ pixels for optimization is not trivial, we compare with an alternative approach in which all input pixels corresponding to the same locations of the predicted high-drop pixels are reduced by 10\% across all layers (all input images corresponding to all the 9 layers). For example, for testcase 7, we scale 1547 pixels in each of 9 input images at the pixels where the high-drop was predicted. As can be seen in the last column of the table (labeled `w/o saliency'), the number of predicted IR drop pixels actually increase on-average. Therefore saliency maps help to meaningfully reduce the number of high-drop pixels and by a significant amount, and using only a tiny portion of the PDN edges.

\begin{table*}[h!]
    \centering
    \small
        \caption{Percentage reduction in the number of predicted high-drop output pixels after optimizing only the top $K$ pixels on the input image corresponding to the highest contributor layer as identified with saliency maps.}
        
    \begin{tabular}{cc|c|c|c|c|c}
    \toprule
         & & \multicolumn{4}{|c|}{With  Saliency Maps} &  \multicolumn{1}{|c}{Without Saliency Maps}
         \\
         \midrule
         &  {\#Highdrop}  & {Layer with}&\multicolumn{1}{c|}{K=100}  &\multicolumn{1}{c|}{K=300}  &\multicolumn{1}{c|}{K=500}  & K=9$\times$\#Highdrop-Pixels \\ 
         & Pixels & Highest Contribution  &\%Reduction & \%Reduction & \%Reduction  & \%Reduction \\
         \midrule
        Testcase7 & 1574 & M1 & 15.00 & 33.30 & 47.25 & 11.07 \\
        Testcase8 & 2502 & M1 & 9.69 & 26.70 & 39.774 & 4.52 \\
        Testcase9 & 550 & M4 & 6.49 & 19.34 & 30.11 & -5.51 \\
        Testcase10 & 364 & M4 & 26.17 & 54.30 &  71.48 & 17.19 \\
        Testcase13 & 211 & M1& 8.98 & 21.56 & 34.13& 1.80 \\
        Testcase14 & 235 & M1& 13.37 & 22.67 & 33.72& 1.74 \\
        Testcase15 & 31 & Via M1M4& 24.47 & 37.23 & 43.62 & -3.19 \\
        Testcase16 & 2464& M1 & 9.49 & 23.85 & 37.40 & -6.91 \\
        Testcase19 & 840 & M8& 42.71 & 82.29 & 86.46 & 36.98 \\
        Testcase20 & 221 & M1& 28.47 & 51.60 & 66.55 & -21.35\\
        \midrule
        Average &&&18.48\% &  37.28\% & 49.05\% & 3.75\% \\
        \bottomrule

    \end{tabular}

    \label{tab:explainability}\vspace{-3mm}
\end{table*}

\begin{figure*}
    \centering
    \includegraphics[width=0.9\linewidth]{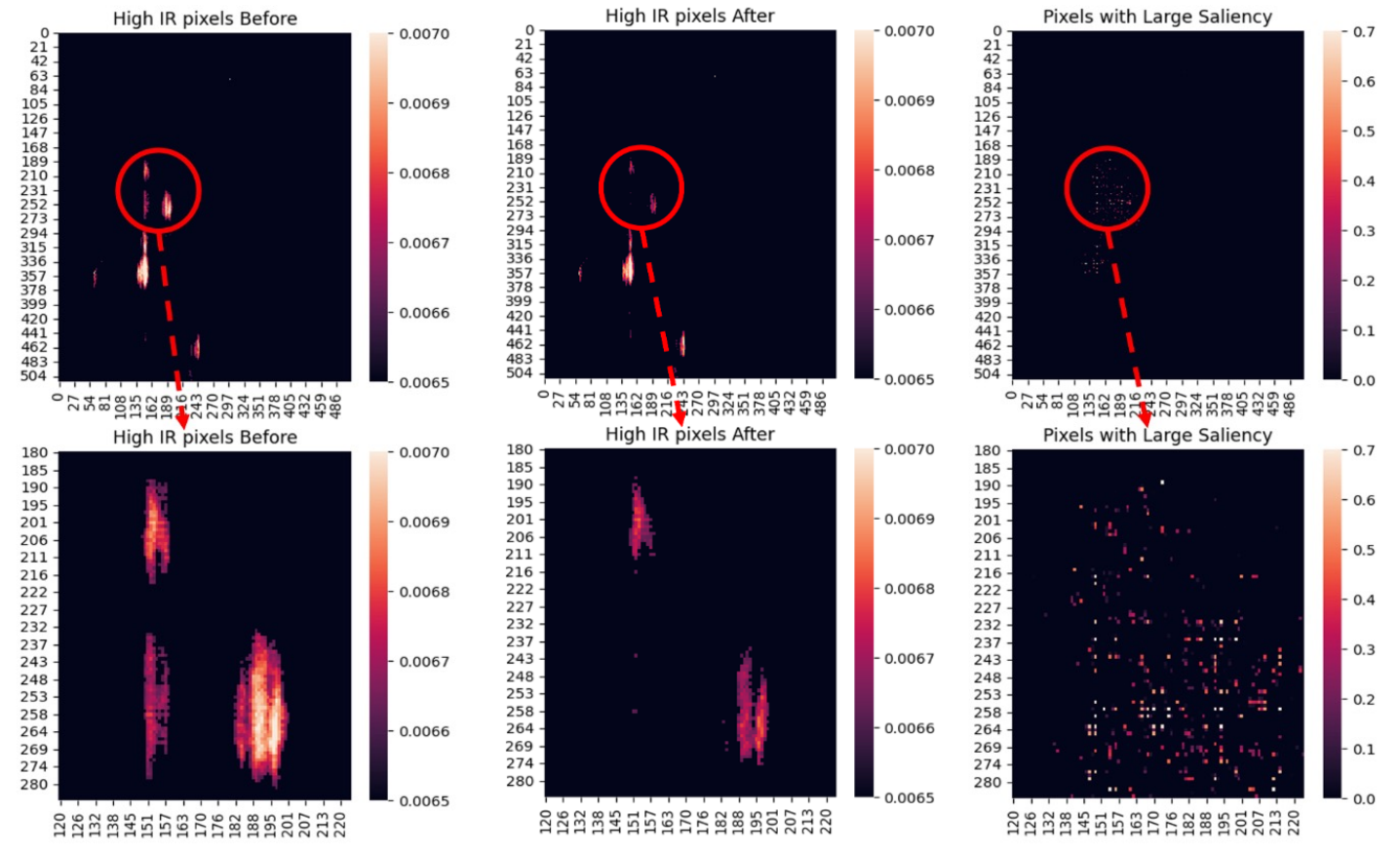}
    \caption{Predicted high-drop pixels before  / after optimization (left / middle)  for testcase 16. High saliency input pixels (right).}\vspace{-3.5mm}
    \label{fig:high-IR}
\end{figure*}

Figure \ref{fig:high-IR} shows the high-drop pixels in the predicted IR-drop map of testcase 16 before and after optimization. The top $K$ pixels with highest saliency on M1 are shown on the right column.


\section{Conclusions}\label{sec:conclusion}

We presented a computer-vision based approach to predicting static IR drops in power delivery networks (PDNs) using the \texttt{AttUNet} model, an advanced variant of the U-Net architecture enhanced with attention mechanisms. Our findings demonstrate that \texttt{AttUNet} significantly surpasses previous models and 2023 ICCAD contest winner in prediction quality. The use of vector concatenation-based attention gates allows \texttt{AttUNet} to selectively prioritize significant areas of feature maps, thus improving the accuracy and reliability of the predictions. This is particularly valuable given the sparse nature of IR drop data and the complex multi-layer structure of modern PDNs. The two-phase training process, leveraging both artificially-generated and real-world data, addresses the challenge of limited real design data and prevents overfitting. Furthermore, the use of saliency maps adds explainability to the model, making it possible to identify the significant causes of predicted high IR drops to help mitigate potential IR-drop violations.

\section{Acknowledgements}

This work is supported by a grant from National Science
Foundation under Award No. 2322713.

\bibliographystyle{unsrt}  
\bibliography{references}  

\begin{thebibliography}{10}

\bibitem{UNet_Sachin_2021}
V.~A. Chhabria, V.~Ahuja, A.~Prabhu, P.~Nikhil, P.~Jain, and S.~S. Sapatnekar.
\newblock Thermal and {IR} drop analysis using convolutional encoder-decoder networks.
\newblock In {\em ASP-DAC}, pages 690--696, 2021.

\bibitem{Multigrid_Kozhaya_2002}
J.~Kozhaya, S.~Nassif, and F.~Najm.
\newblock A multigrid-like technique for power grid analysis.
\newblock In {\em TCAD}, volume~21, pages 1148--1160, 2002.

\bibitem{FastAlg_Zhong_2005}
Y.~Zhong and M.~D.~F. Wong.
\newblock Fast algorithms for {IR} drop analysis in large power grid.
\newblock In {\em ICCAD}, pages 351--357, 2005.

\bibitem{locality}
S.~Köse and E.~G. Friedman.
\newblock Fast algorithms for {IR} voltage drop analysis exploiting locality.
\newblock In {\em DAC}, pages 996--1001, 2011.

\bibitem{Hierarchical}
M.~Zhao, R.~V. Panda, S.~S. Sapatnekar, T.~Edwards, R.~Chaudhry, and D.~Blaauw.
\newblock Hierarchical analysis of power distribution networks.
\newblock In {\em TCAD}, volume~21, pages 159--168, 2002.

\bibitem{Krylov_Hao_2001}
T.~Chen and C.~C.-P. Chen.
\newblock Efficient large-scale power grid analysis based on preconditioned {Krylov}-subspace iterative methods.
\newblock In {\em DAC}, pages 559--562, 2001.

\bibitem{Krylov_Chou_2011}
C.~Chou, N.~Tsai, H.~Yu, C.~Lee, Y.~Shi, and S.~Chang.
\newblock On the preconditioner of conjugate gradient method — a power grid simulation perspective.
\newblock In {\em ICCAD}, pages 494--497, 2011.

\bibitem{ECO-based_1}
S.~Lin, Y.~Fang, Y.~Li, Y.~Liu, T.~Yang, S.~Lin, C.~Li, and E.~Fang.
\newblock {IR} drop prediction of {ECO}-revised circuits using machine learning.
\newblock In {\em VTS}, pages 1--6, 2018.

\bibitem{ECO-based_2}
S.~Kundu, M.~Prasad, S.~Nishad, S.~Nachireddy, and K.~Harikrishnan.
\newblock {MLIR}: Machine learning based {IR} drop prediction on {ECO} revised design for faster convergence.
\newblock In {\em VLSID}, pages 68--73, 2022.

\bibitem{PowerNet}
Z.~Xie, H.~Ren, B.~Khailany, Y.~Sheng, S.~Santosh, J.~Hu, and Y.~Chen.
\newblock {PowerNet}: Transferable dynamic {IR} drop estimation via maximum convolutional neural network.
\newblock In {\em ASP-DAC}, pages 13--18, 2020.

\bibitem{BeGAN_Chhabria_2021}
V.~A. Chhabria, K.~Kunal, M.~Zabihi, and S.~S. Sapatnekar.
\newblock {BeGAN}: Power grid benchmark generation using a process-portable {GAN}-based methodology.
\newblock In {\em ICCAD}, pages 1--8, 2021.

\bibitem{ICCAD2023-contest}
{CAD Contest at ICCAD}.
\newblock \url{https://drive.google.com/file/d/162C8PI1umxad3uYrO6aBYgCmjeTQbvR4/view}, 2023.

\bibitem{Saliency_Simonyan_2014}
K.~Simonyan, A.~Vedaldi, and A.~Zisserman.
\newblock Deep inside convolutional networks: Visualising image classification models and saliency maps.
\newblock In {\em ICLR}, 2014.

\bibitem{CNNExplainer_Zhang_2018}
Q.~Zhang, Y.~Yang, Y.~Liu, Y.~Wu, and S.~Zhu.
\newblock Unsupervised learning of neural networks to explain neural networks.
\newblock {\em arXiv: 1805.07468}, 2018.

\bibitem{XCNN_Tabanaei_2020}
A.~Tavanaei.
\newblock Embedded encoder-decoder in convolutional networks towards explainable {AI}.
\newblock {\em arXiv: 2007.06712}, 2020.

\bibitem{DAttn_Seo_2017}
S.~Seo, J.~Huang, H.~Yang, and Y.~Liu.
\newblock Interpretable convolutional neural networks with dual local and global attention for review rating prediction.
\newblock In {\em RecSys}, pages 297--305, 2017.

\bibitem{AttUNet}
O.~Oktay, J.~Schlemper, L.L. Folgoc, M.~Lee, M.~Heinrich, K.~Misawa, K.~Mori, S.~McDonagh, B.~Kainz N.~Y.~Hammerla, and B.~Glockerand~D. Rueckert.
\newblock {Attention U-Net}: Learning where to look for the pancreas.
\newblock {\em arXiv: 1804.03999}, 2018.

\bibitem{Wang_NonLocalNN_2017}
X.Wang, R.~B. Girshick, A.~Gupta, and K.~He.
\newblock Non-local neural networks.
\newblock {\em arXiv: 1711.07971}, 2017.

\bibitem{ImageNet_Krizhevsku_2012}
A.~Krizhevsky, I.~Sutskever, and G.~E. Hinton.
\newblock {ImageNet} classification with deep convolutional neural networks.
\newblock In {\em NeurIPS}, page 1097–1105, 2012.

\bibitem{cosine_anneal}
I.~Loshchilov and F.~Hutter.
\newblock {SGDR}: Stochastic gradient descent with warm restarts.
\newblock In {\em ICLR}, 2017.

\end{thebibliography}






\end{document}